\documentclass[amstex,amssymb,12pt]{article}
\usepackage{amsmath}
\usepackage{amsfonts}
\usepackage{amssymb}
\usepackage{graphicx}
\begin{document}

\author{G. Musulmanbekov, A. Al--Haidary}
\title{Fragmentation of Nuclei at Intermediate and High Energies in Modified
Cascade Model }
\date{}
\maketitle

\begin{abstract}
The process of nuclear multifragmentation has been implemented, together
with evaporation and fission channels of the disintegration of excited
remnants in nucleus-nucleus collisions using percolation theory and the
intranuclear cascade model. Colliding nuclei are treated as
face--centered--cubic lattices with nucleons occupying the nodes of the
lattice. The site--bond percolation model is used. The code can be applied
for calculation of the fragmentation of nuclei in spallation and
multifragmentation reactions.
\end{abstract}

%\medskip

\section{Introduction}

The intranuclear cascade model is one of the basic tools for analyzing
spallation and multifragmentation processes in nuclear collisions. In the
traditional cascade model of hadron-nucleus and nucleus-nucleus
interactions, particle production is treated in two stages. In the first
fast stage, intranuclear cascade occurs inside the target and (or) the
projectile nuclei and some nucleons from the target and projectile nuclei
are knocked out, together with mesons. In the second stage, residual
nuclei (generally, in an excited state) divide into two remnants in the
fission channel or evaporate protons, neutrons and/or light nuclei,
including helium isotopes. However, experimental data indicate that, at
intermediate energies, a third competing process, multifragmentation,
comes into play, in which excited remnants break up into intermediate mass
fragments (IMF). There are two approaches for theoretical description of
multifragmentation: dynamical and statistical. In statistical
multifragmentation models, an excited remnant achieves thermal equilibrium
state and then expands, eventually reaching the freeze--out volume. At
this point it fragments into neutrons, light charged particles and IMFs.
In dynamical models IMFs are formed at the fast stage of nuclear collision
via dynamical forces between nucleons during the evolution of the total
system of interacting projectile and target. In this case the whole system
and its parts (projectile and target remnants) never pass through states
of thermal equilibrium.

There is one more approach for description of the process of
multifragmentation: percolation theory. Percolation models treat the
nucleus as a lattice with nucleons located at nodes of the lattice. It has
been found that results of percolation calculations depend significantly
upon the details of the lattice structure. For reasons of computational
convenience, the simple cubic lattice has been most frequently used in
multifragmentation simulations\cite{bauer}, but several studies have found
\cite{chao, santiago} that the face-centered-cubic lattice more accurately
reproduces the experimental distributions of fragment masses and their
energy spectra.

Although lattice simulations have been found to reproduce
multifragmentation data surprisingly well, there has been little
examination of the role of the lattice arrangement of nucleons inside
nuclei. That is, lattices were employed more as computational techniques,
rather than as formal nuclear models. The appearance of solid state models
of nuclear structure can be dated from the paper of Pauling in 1965
\cite{pauling, astatos, cook1}.  The most attractive lattice model is the
face--centered--cubic (FCC) model proposed by Cook and
Dallacasa\cite{cook1} because it brings together shell, liquid-drop and
cluster characteristics, as found in the conventional models, within a
single theoretical framework. Unique among the lattice models, the FCC
reproduces the entire sequence of allowed nucleon states as found in the
shell model.

In the present paper we further develop the modified intranuclear
cascade--evaporation code, (MCAS) elaborated by one of the authors \cite
{mcas}, with the aim of inclusion of multifragmentation channels. The word
''modified'' relates to the implementation of the concept of "formation
time" into traditional cascade calculations, as described in Section 3.
The goodness of fit of the MCAS to experimental data has been reported in
previous papers \cite{mcas, krasnov} concerning multiparticle production
in nucleus-nucleus collisions at intermediate and moderately high energies
(up to 10--20 GeV/n). Since traditional cascade models consider nuclear
structure as a dilute fermi gas, we reconstructed the model in the
framework of the lattice nuclear model. For this purpose we implemented
the FCC lattice arrangement of nucleons for the colliding nuclei according
to the algorithm proposed in reference \cite{cook2} (Section 2).
Calculations of multifragmentation channels are performed on the basis of
the bond-site model of percolation theory (Section 4). Comparisons with
experimental data are given in Section 5.

\section{FCC\ lattice model of nuclear structure}

The FCC packing of nucleons, with protons and neutrons occupying lattice
sites in alternating layers, can be seen as consisting of four
interpenetrating cubes. A nearest-neighbor distance of about 2.0262 fm
reproduces the known core density of nuclei (0.17 nucleons/fm$^{3}$). The
essence of the geometry of the FCC model can be shown using the quantum
numbers that are assigned to each nucleon in the conventional shell model
\cite{cook2}. It is known that a nucleon's distance from the center of the
nucleus determines its principal quantum number {\bf n. }The distance of
the nucleon from the ''nuclear spin axis'' determines its total angular
momentum (quantum number{\bf \ j}). Finally, the distance of each nucleon
from the $y-z$ plane determines its magnetic quantum number {\bf m}. The
inherent simplicity of the FCC model is evident in the FCC definitions of
the eigenvalues:

\begin{eqnarray}
{\bf n} &=&(\left| x\right| +\left| y\right| +\left| z\right| -3)/2, \\
{\bf j} &=&(\left| x\right| +\left| y\right| -1)/2, \\
\left| {\bf m}\right| &=&\left| x\right| /2,
\end{eqnarray}
where the sign of the ${\bf m}$ value is determined by the intrinsic spin
orientation of the nucleon in the antiferromagnetic lattice (spin up
=$\frac{1}{2}$ and spin down =--$\frac{1}{2}$). Conversely, the coordinate
values can be determined solely from the nucleon eigenvalues:
\begin{eqnarray}
x &=&\left| 2m\right| (-1)^{m+\frac{1}{2}}, \\
y &=&(2j+1-\left| x\right| )(-1)^{i+j+m+\frac{1}{2}}, \\
z &=&(2n+3-\left| x\right| -\left| y\right| )(-1)^{i+n-j-1},
\end{eqnarray}
where ${\bf i}$ is the isospin quantum number. Therefore, knowing the full
set of eigenvalues for a given set of nucleons, the configuration of those
nucleons in 3--D space relative to the nuclear center can be determined
unambiguously. Using the fermi coordinates of each nucleon, the mean
radius of the nucleus with A nucleons is defined as
\begin{equation}
R\left[ A\right] =R_{nucleon}+\frac{1}{A}\stackrel{A}{\sum }r_{j},
\end{equation}
where $r$ is the Euclidean distance of each nucleon, $\sqrt{%
x_{j}^{2}+y_{j}^{2}+z_{j}^{2}},$ from the origin and $R_{nucleon}$ is the
nucleon radius. The calculated charged radii for various nuclei are in
good agreement with experiment.

\section{Intranuclear Cascade with the Nuclear Lattice Model}

Nucleon coordinates (4 -- 6) for the target (projectile) nucleus are
generated in accordance with the algorithm given in reference
\cite{cook3}. For each nuclear collision, lattices of target and
projectile nuclei are oriented randomly in relation to the collision axes.
This random orientation of the nuclear lattice in 3-D space mimics the
Woods-Saxon distribution of nuclear density for medium and heavy nuclei.
Nucleon momenta inside the nucleus, ${\bf p}${\bf ,} are generated
uniformly in the space $0\leq \left| {\bf p}\right| \leq p_{F} $. The
bound Fermi momentum $p_{F}$ relates to the local nucleon density as
\begin{equation}
p_{F}=(3\pi ^{2})^{1/3}h\rho ^{1/3}(r).
\end{equation}

An inelastic collision of two nuclei is an incoherent superposition of
baryon--baryon, meson--baryon and meson--meson elastic and inelastic
interactions. Elastic and inelastic cross sections and kinematical
features of the elastic scattering are taken from experiments. Description
of the inelastic event generator is given in Appendix. All interactions
are arranged into four groups.\newline Group C -- interactions of the
nucleons of the projectile nucleus with those from the target nucleus. All
secondary particles produced in any group of interactions are considered
as cascade particles.\newline Group A -- interactions of the cascade
particles with the nucleons of the target nucleus;\newline Group B --
interactions of the cascade particles with the nucleons of the incident
nucleus;\newline Group D -- so called ''cascade--cascade'' interactions
--- interactions of cascade particles with each other.

The probability of interaction of particles $i$ and $j$ is defined by a
black disk approximation:
\begin{equation}
P(b_{ij}^{2})=\Theta (b_{ij}^{2}-\sigma _{tot}/\pi ),
\end{equation}
where $b_{ij}$ is the impact parameter between hadrons $i$ and $j$, and
$\sigma_{tot} $ is their total cross section. Cross sections of resonances
in subsequent interactions are taken to be the same as for stable
particles. The evolution of the interacting system is considered as
follows. At some instant of time $t$ all possible interacting pairs in
each group (A,B,C,D) are determined. Among all possible interactions that
one is chosen to be the first if it occurs before others, i.e., $\Delta
t=min\{t_{i}\}$; then the positions of both nuclei and all cascade
particles are moved to new positions corresponding to a new instant of
time $t_{i}\rightarrow t_{i}+\Delta t$. Since the formation of hadronic
states of secondary particles takes some time, we apply the concept of
formation time (zone) for consideration of their subsequent interactions.
The formation time relates to the development of the cross section of the
produced particle during its propagation inside the nuclear medium. We use
the exponential form of the evolution of cross sections until the
subsequent collision occurs
\begin{equation}
\sigma _{2}^{l}=\sigma _{1}^{I}(1-(1-x^{l})e^{-\tau /(\gamma \tau _{0})})
\end{equation}
for the leading particle,
\begin{equation}
\sigma _{2}^{m}=\sigma _{1}^{m}-(\sigma _{1}^{m}-x^{m}\sigma
_{1}^{I})e^{-\tau _{1}/(\gamma \tau _{0})}
\end{equation}
for the $m$-th produced particle, where ${\sigma _{1}^{I}}$ is the normal
cross section of the incident particle in the first collision, ${\sigma
_{2}^{l}}$ is the cross section of the remnant of the projectile (leading
particle) in the second collision, $\sigma _{2}^{m}$ is the cross section
of the $m$-th produced particle, $\sigma _{1}^{m}$ is the cross section
for this type of particle in the normal state, $\gamma $ is the Lorentz-
factor and $\tau _{0}$ is an adjustable parameter corresponding to the
mean value of the formation time in the rest frame of the particle. For
${r+1}$-th inelastic rescattering of the incident particle, the cross
section is defined as
\begin{equation}
\sigma _{r+1}^{l}=\sigma _{1}^{I}\prod_{i=1}^{r}(1-(1-x_{i}^{l})e^{-\tau
_{i}/(\gamma _{i}\tau _{0})})
\end{equation}

Among secondaries, s--wave resonances (deltas, rho and omega mesons) can
be produced. The hadronic event generator is briefly described in the
Appendix. During the evolution of the system, the produced resonances may
decay before their subsequent interactions. A check is made whether the
Pauli principle is satisfied both for all interactions and for the decay
of resonances. The cascade stage of particle generation is completed when
all cascade particles have left both nuclei or have been partly absorbed
by them. In this way, the first fast stage of multiparticle production of
the nuclear collision has been completed. After replacing the fermi--gas
nuclear model by the FCC lattice, we compared the results of simulations
using both models on multiparticle production in intermediate and high
energy nuclear collisions and have found that they are identical. The
first measurable characteristic of nucleus-nucleus collisions is the
reaction cross section. In intranuclear cascade models the cross section
is defined by the ratio of the number of realized inelastic collisions,
$N^{in}$, to the total number of trials, $N^{trial}$:
\begin{equation}
\sigma ^{reac}=\frac{N^{in}}{N^{trial}}\pi (R_{A}+R_{B}+\Delta )^{2},
\end{equation}
where $R_{A,B}$ is the radii of the  colliding nuclei and $\Delta $ is the
radius of the strong interaction. The fermi--gas and FCC lattice models
are in agreement to within an accuracy of 5 percent.

\section{Fragmentation of Excited Remnants}

The number and total charge of the remaining nucleons in each remnant
specify the mass and charge numbers of the residual nuclei. In general,
remnants are in excited states and possess angular momentum. The
excitation energy of each remnant nucleus is determined by the energy of
the absorbed particles and the ''holes''\ remaining after nucleons have
been knocked out during the intranuclear cascade process. The momentum and
angular momentum of the residual nucleus are evaluated in light of the
conservation of momentum and sequentially followed for each intranuclear
interaction. Thus, there are three competing processes for the
disintegration of the excited remnant nucleus: evaporation, fission and
multifragmentation. In the standard intranuclear cascade model, only the
first two processes are taken into account \cite{barash}. The purpose of
the present study was to implement multifragmentation on the basis of
percolation theory and to determine the relative weights of the above
three competing processes. We have done this by applying the site-bond
percolation model. We assume that nucleons occupying lattice sites are
connected with their neighbors via bonds which schematically represent two
body nuclear forces. In the fast stage, during the development of
intranuclear cascade some nucleons occupying the sites of the FCC lattice
of the target (projectile) nucleus are knocked out, leaving ''holes'' at
those sites. We say that these sites are broken. The ratio of the number
of broken sites to the total number of sites (the mass number of the
target or projectile) characterizes the degree of destruction of the
target (projectile) nucleus after the cascade stage. This ratio depends on
the collision energy, the mass numbers of colliding nuclei and,
particularly, on the impact parameter of the collision. In peripheral
collisions, mainly peripheral nucleons are knocked out, meaning that, with
high probability, the remaining nucleons form one cluster in which all
sites are occupied. In collisions with more centrality, corresponding to
intermediate or small impact parameters, nucleons are knocked out mainly
from the nuclear interior and the target (projectile) remnant represents
the lattice with some sites broken. As mentioned above, remnants, in
general, are in excited states. The larger the impact parameter, the
smaller is the number of broken sites and the less is the excitation
energy of the remnant. This initial condition is preferable for
equilibration and thermalization of excited nuclear media and allows one
to use evaporation and fission mechanisms for subsequent disintegration of
the excited remnant. With increasing centrality of the collision, the
number of broken sites increases (large destruction), leading to
increasing excitation energy of the remnant nucleus. For this case, there
is no conventional understanding of the mechanism of disintegration of an
excited remnant (thermal break-up with statistical multifragmentation,
liquid-gas phase transition, sequential evaporation, cold shattering
break-up, etc.). However, it is obvious that when there is considerable
destruction of the remnant, there is no possibility for equilibration and
thermalization over the whole volume of the remnant.

For excited remnant disintegration we specify the bond breaking
probability, as an input parameter $p_{bond},$ in the form of impact
parameter dependence:
\begin{equation}
p_{bond}(b)=p_{bond}(0)\sqrt{1-\frac{b^{2}}{(R_{A}+R_{B})^{2}},}
\end{equation}
where $R_{A}$ and $R_{B}$ are the radii of the colliding nuclei. This
anzatz can be derived from considerations of the collision geometry. The
cluster counting algorithm, developed by authors, looks for clusters
(fragments): whether neighboring nucleons are connected via bonds or not.
Only first--nearest and second--nearest neighbors are taken into account
in the counting algorithm. In the initial FCC lattice each nucleon has 12
first--nearest neighbors at a distance of 2.0262 fm and 6 second--nearest
neighbors at 2.8655 fm. As a result of this counting algorithm, we obtain
the mass and charge distribution of the fragments. Although this approach
is statistical and the probability of any bond to be broken does not
depend on its position, the probability of disintegration of the remnant
on multiple clusters (fragments) will be higher in the vicinity of regions
with many broken sites. From this it follows that the process of
multifragmentation is influenced by the dynamics of the collision, i.e.,
according to our scenario it is not purely a statistical process.

Next we specify the energetic characteristics of the radiated fragments.
In general, in its proper frame the remnant possesses rotational energy,$
E^{rot}$, and excitation energy, $E^{\ast }$, which are used in the
summation of the rotational, $E_{fr}^{rot}$, and kinetic energies of fthe
ragments, $ E_{fr}^{kin}$, their excitation energies, $E_{fr}^{\ast }$,
and  the energy of the coulomb interactions of the fragments,
$E_{fr}^{coul}$:
\begin{multline}
E^{rot}+E^{\ast }=E_{fr}^{rot}+E_{fr}^{kin}+E_{fr}^{\ast }+E_{fr}^{coul}=
\\ \sum E_{i}^{rot}+\sum E^{kin}(A_{i},Z_{i})+\sum E^{\ast
}(A_{i},Z_{i})+\frac{ 1}{2}\sum \frac{Z_{i}Z_{j}}{r_{ij}}.
\end{multline}
In the standard intranuclear cascade model the contribution of rotational
energy, $E_{fr}^{rot},$ is small compared with other terms, at least for
light nuclei, as projectiles. Whether or not this is the case in reality
is unknown. Large rotational energies could be realized in this approach
if we included nuclear viscosity. In the current calculations we neglect
the first term. Moreover, for computational convenience we make additional
simplifications in Eq. (15). Since the coulomb repulsion of the charged
fragments increases their kinetic energies, we define the resulting
kinetic energies of the fragments as follows:
\begin{equation}
E_{fr}^{kin}+E_{fr}^{coul}=\sum E^{kin}(A_{i},Z_{i})+\frac{1}{2}\sum
\frac{ Z_{i}Z_{j}}{r_{ij}}=\sum \varepsilon _{i}(A_{i},Z_{i}).
\end{equation}
Another simplification concerns the excitation of the fragments: we assume
that only one fragment among others is excited, the mass number of which
is maximal. It is justified, particularly, when comparing model with data
obtained through inverse kinematics because the experimental setup
registers a majority of radioactive fragments as well. Therefore, the
excitation energy of the remnant, $E^{\ast }$, is converted into the
kinetic energies of the fragments and the excitation energy of the
fragment with maximal mass:
\begin{equation}
E^{\ast }=\sum \varepsilon _{i}(A_{i},Z_{i})+E^{\ast }(A_{\max },Z_{\max })
\end{equation}

With these simplifications we generate the energy distribution of
fragments applying considerations proposed in reference \cite{campi85}.
Before the collision the nucleons have a momentum distribution that is
uniform inside the Fermi sphere of radius $p_{F}.$ After the collision the
distribution in the vicinity of the beam propagation is wider because of
intranuclear interactions accompanied by local excitation of the nuclear
medium. This can be written in the form:
\begin{equation}
n(\varepsilon )\varpropto 1/\left\{ 1+\exp \left[ \frac{\varepsilon
-\varepsilon _{F}}{T_{eff}}\right] \right\} ,
\end{equation}
where $\varepsilon =p^{2}/2m$, and $\varepsilon _{F}$ is the boundary
Fermi energy. The ''effective temperature'' is given by
\begin{equation}
T_{eff}=cE^{\ast }/N_{br},
\end{equation}
where $c$ is an adjustable parameter and $N_{br}$ is the number of broken
sites. Kinetic energies of nucleons composing the fragment are generated
according to distribution (18), and, summing up all vector momenta
directed randomly in 3-D space, we obtain the momentum of the fragment. In
such a way, we generate momenta of all produced fragments. The remaining
part of the remnant excitation energy, (17), is assigned to the fragment
with maximal mass number. And, of course, we take into account the
conservation of energy and momentum for the whole reaction.

\section{Comparison with Experiment}

Observation of residues emerging from spallation reactions in direct
kinematics still remains a difficult task.
\begin{figure}[ftbp]
\begin{center}
\includegraphics[width=3.6in]{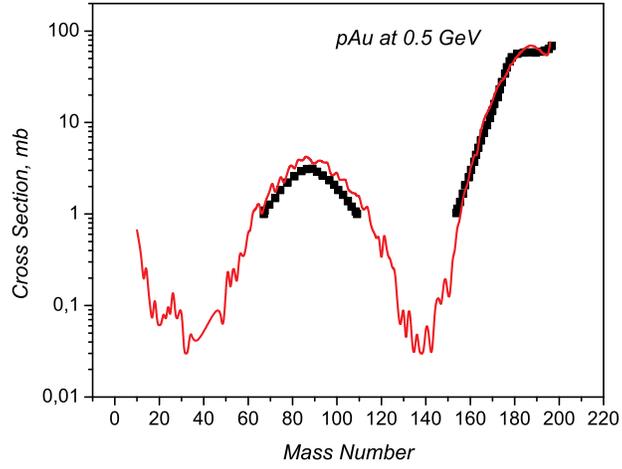}
\caption{Mass distribution of residues produced in 0.5 GeV proton induced
reaction on $^{197}$Au; $p_{bond}=0.5$. Data are from
paper\protect\cite{pAu-ms05}.}
\end{center}
\end{figure}
\begin{figure}[ftbp]
\begin{center}
\includegraphics[width=3.6in]{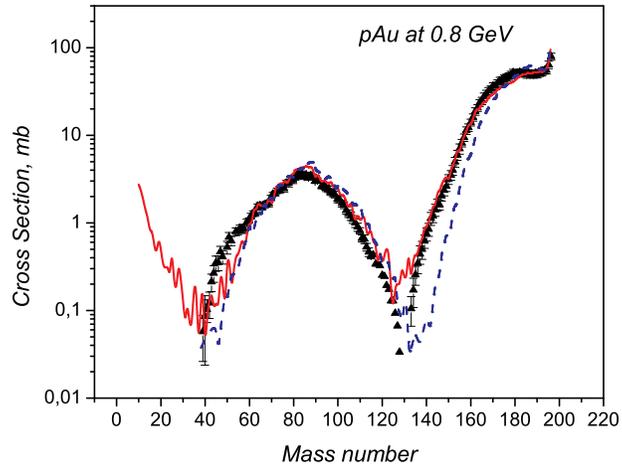}
\caption{Mass distribution of residues in the reaction $^{197}$Au + p at
800 A MeV; $p_{bond}=0.57$. Dashed curve is outcome of calculations
without the contribution of multifragmentation channels . Data are taken
from paper\protect\cite{pAuneut}.}
\end{center}
\end{figure}
\begin{figure}[ftbp]
\begin{center}
\includegraphics[width=3.4in]{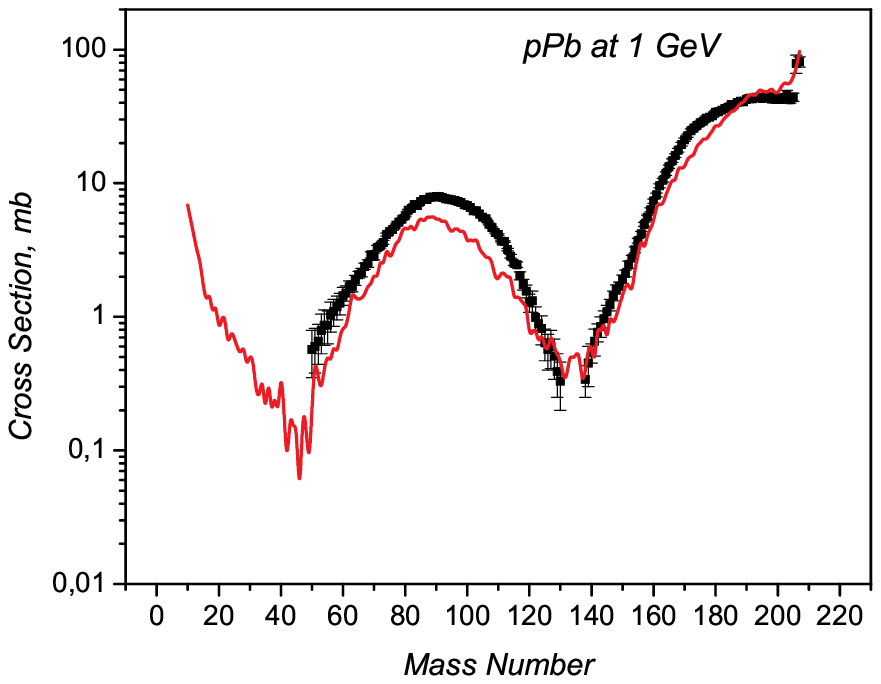}
\caption{Mass distribution of residues in the reaction $%
^{208}$Pb + p at 1 A GeV; $p_{bond}=0.6$. Data are taken from
paper\protect\cite{pPbneut}.}
\end{center}
\end{figure}
\begin{figure}[ftbp]
\begin{center}
\includegraphics[width=3.4in]{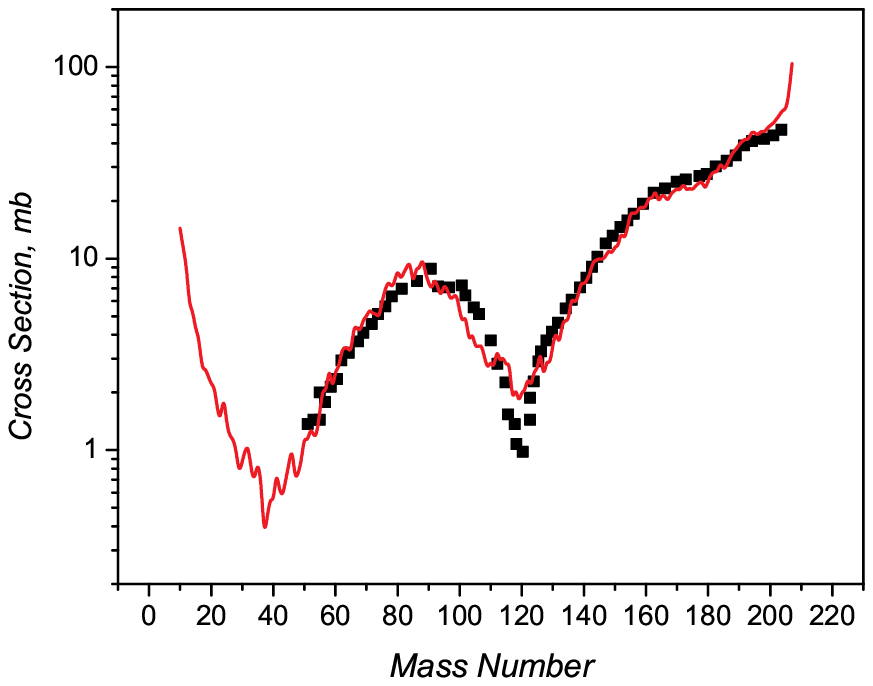}
\caption{Mass distribution of residues in the reaction $%
^{208}$Pb + d at 1 A GeV; $p_{bond}=0.62$. Data are taken from
paper\protect\cite{dPbneut}.}
\end{center}
\end{figure}
Collisions of protons and light nuclei with heavy ions performed at GSI in
inverse kinematics allows one to determine the production of residues
prior to $\beta $ decay. This provides a good opportunity to compare the
available data with theoretical models to achieve a better understanding
of the mechanisms of reactions which is, today, far from satisfactory.
Until now calculations have been performed by different versions of
intranuclear cascade followed by the evaporation model.
\begin{figure}[ftbp]
\begin{center}
\includegraphics[width=6in]{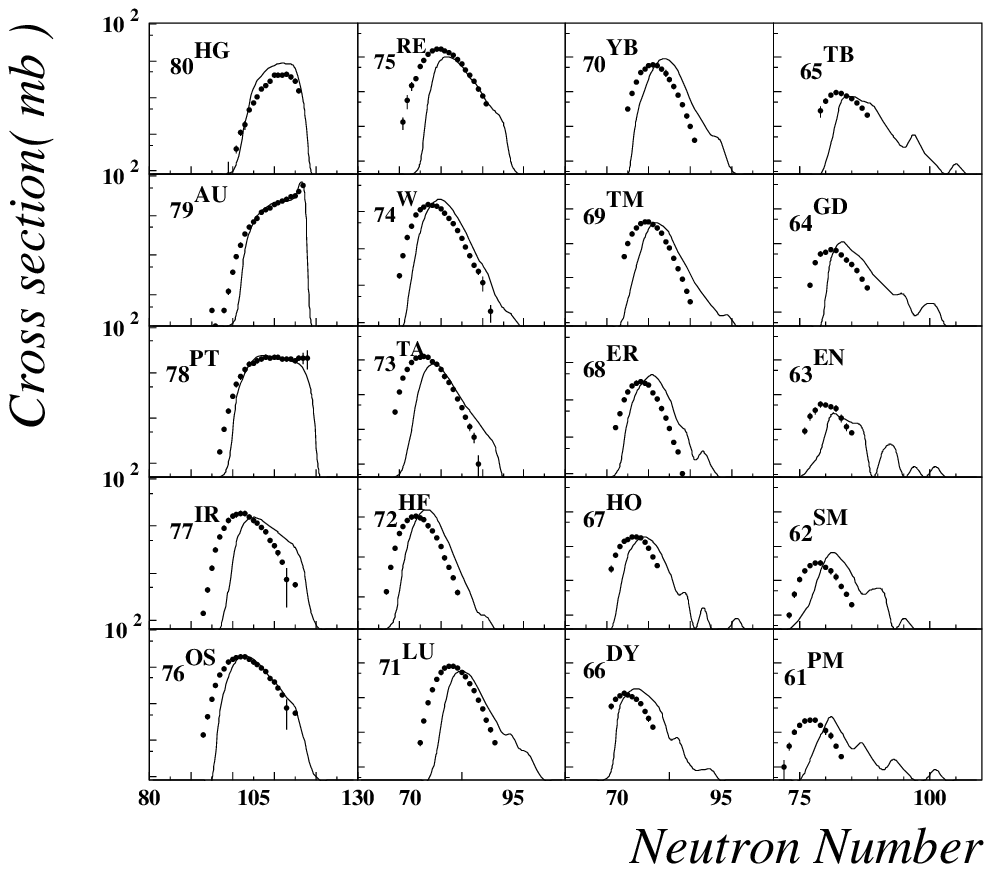}
\caption{Isotopic distribution of spallation residues in reaction
$^{197}$Au + p at 800 A MeV; $p_{bond}=0.57$. Data are taken from the
paper\protect\cite{pAuneut}.}
\end{center}
\end{figure}
\begin{figure}[ftbp]
\begin{center}
\includegraphics[width=6in]{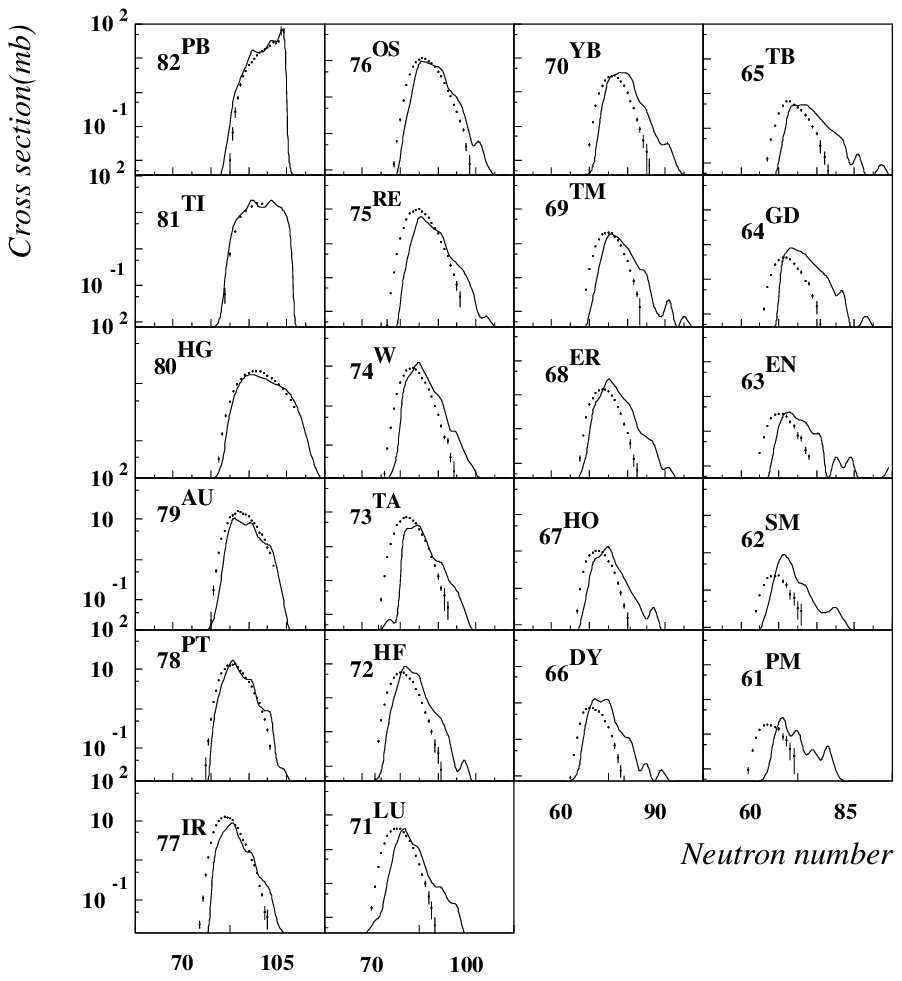}
\caption{Isotopic distribution of spallation residues in reaction
$^{208}$Pb + p at 1 A GeV; $p_{bond}=0.6$. Data are from
paper\protect\cite {pPbneut}.}
\end{center}
\end{figure}
\begin{figure}[ftbp]
\begin{center}
\includegraphics[width=6in]{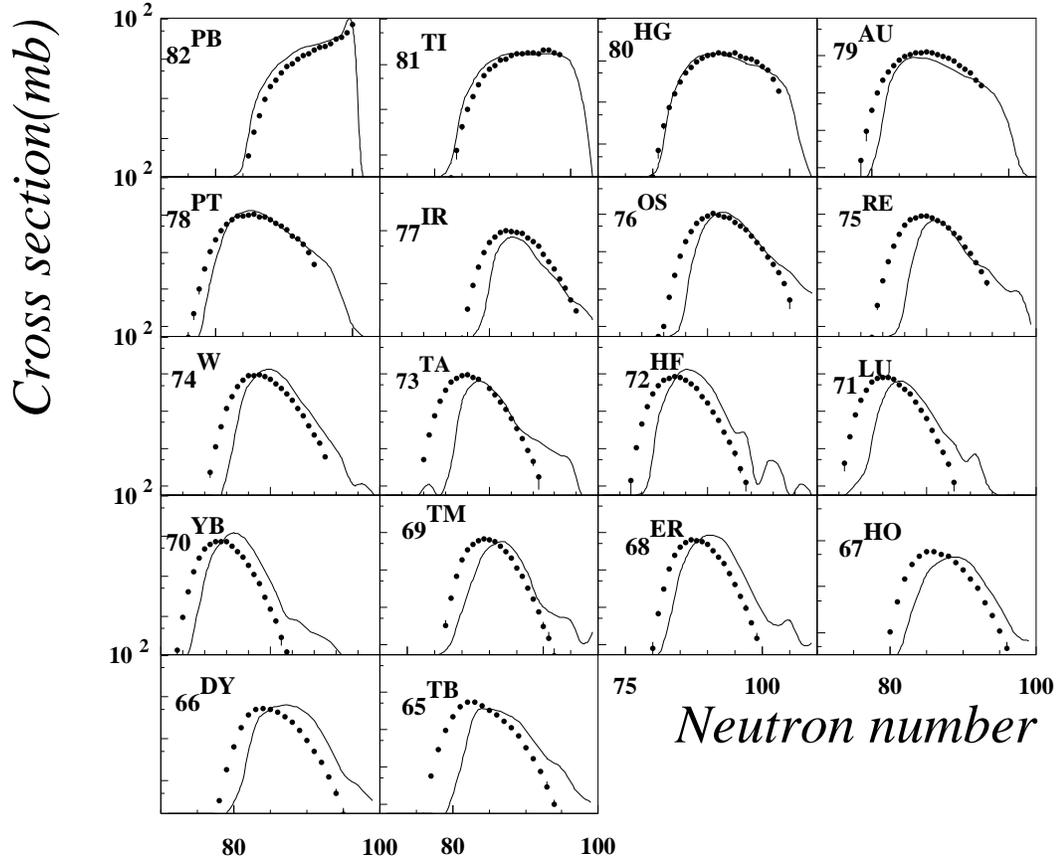}
\caption{Isotopic distribution of spallation residues in reaction
$^{208}$Pb + d at 1 A GeV; $p_{bond}=0.62$. Data are from the
paper\protect\cite{dPbneut}.}
\end{center}
\end{figure}
As seen from the previous section, our model also includes
multifragmentation channels in the framework of the percolation approach.
Here we define the values of the input parameters, which are the bond
breaking probability, $p_{bond}(0)$, in Eq. (14) and the constant $c$ in
Eq. (19). $p_{bond}(0)$ depends on the energy of the collision and the
type of reaction. For a specific reaction, it is obvious that site and
bond breaking probabilities are small at low energies and start growing
with increasing energy, reaching constant values at the regime called
''limiting fragmentation''. Limiting fragmentation is reached at different
energies for different reactions. Therefore, at low energies the
dominating mechanisms of disintegration of excited remnants are
evaporation and fission. As the energy of the collision grows, the
contribution of multifragmentation processes increases, depending on the
site and bond breaking probabilities. Since the number of broken sites is
defined automatically during the development of the intranuclear cascade,
only the bond breaking probability remains to be input as a parameter. For
proton induced reactions, $p_{bond}(0)$ changes from 0, at an incident
proton energy of a few tens of MeV, to 0.77 at the limiting fragmentation
energy (3 -- 4 GeV). With regard to the parameter $c$ in Eq. (19), its
value is chosen to be 0.7, energy independent, for all types of reactions.
Comparison of the model calculations for mass distributions in spallation
reactions pAu at 0.5 and 0.8 GeV, pPb at 1 GeV, and dPb at 2 GeV are shown
in Figs. 1 -- 4. As seen from the figures, at energies lower than those
corresponding to the limiting fragmentation regime, mass yield
distributions of residues in proton-induced reactions have well-pronounced
bell-shaped curves in the central part, corresponding to the contribution
of fission channels. Evaporation channels give a dominating contribution
in the right peak of the distribution with a plateau at high mass
residues.  The values of the level density parameters for evaporation and
fission are taken to be 0.1A MeV$^{-1}$, the same for both and independent
of the type of reaction and collision energy. Figures 5--7 show the
isotopic distributions of residues produced in the following reactions:
pAu at 0.8 GeV, pPb at 1 GeV and deuteron -- Pb at 2 GeV. Calculated
distributions are shifted toward neutron-rich\ isotopes for lighter
residues. We think that underestimation of the proton--rich isotopes and
overestimation of the neutron--rich isotopes could be corrected by better
description of proton--neutron competition in fission--evaporation
channels.
\begin{figure}[ftbp]
\begin{center}
\includegraphics[width=3.4in]{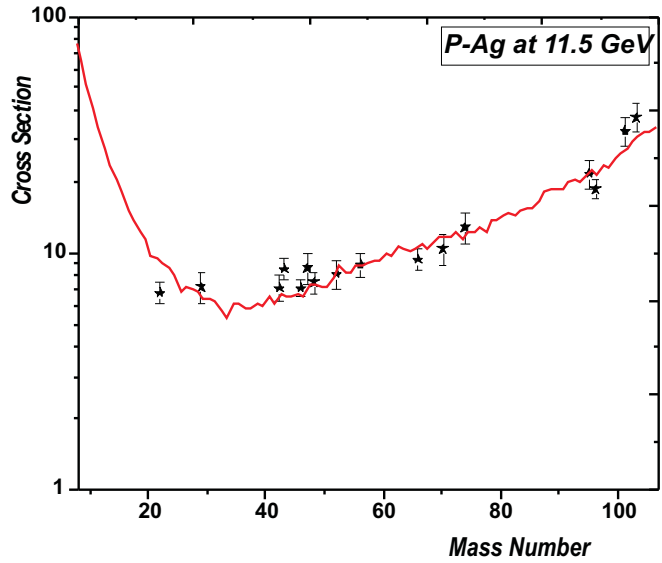}
\caption{Mass distribution of fragments in the reaction p + $^{108}$Ag at
11.5 GeV; $p_{bond}=0.77$. Data are taken from paper \protect\cite{engl}.}
\end{center}
\end{figure}
\begin{figure}[ftbp]
\begin{center}
\includegraphics[width=3.8in]{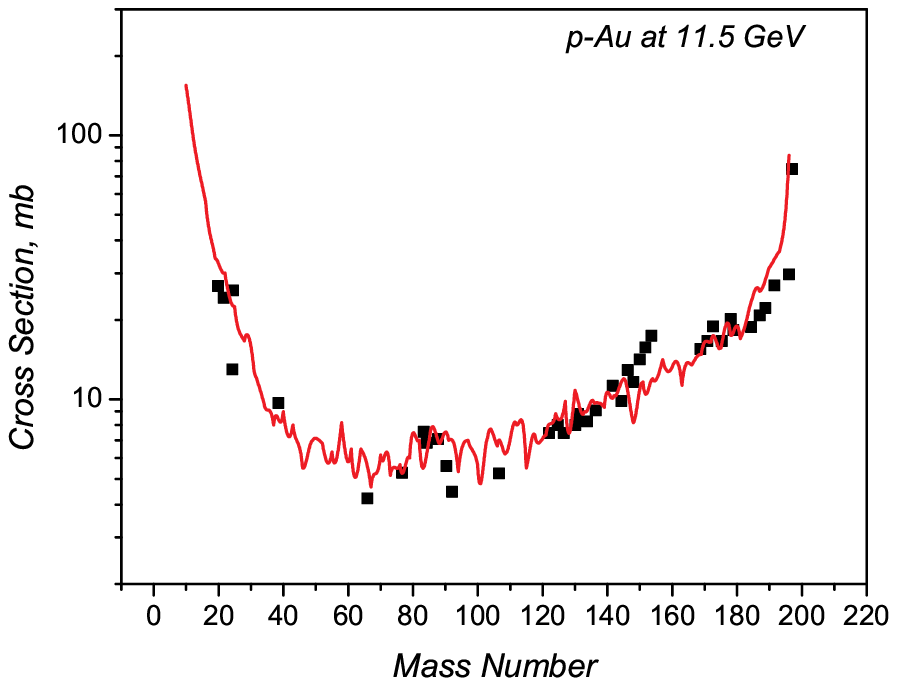}
\caption{Mass distribution of fragments in the reaction p + $^{197}$Au at
11.5 GeV; $p_{bond}=0.77$. Data are taken from paper \protect\cite{kauf}.}
\end{center}
\end{figure}
\begin{figure}[ftbp]
\begin{center}
\includegraphics[width=3.4in]{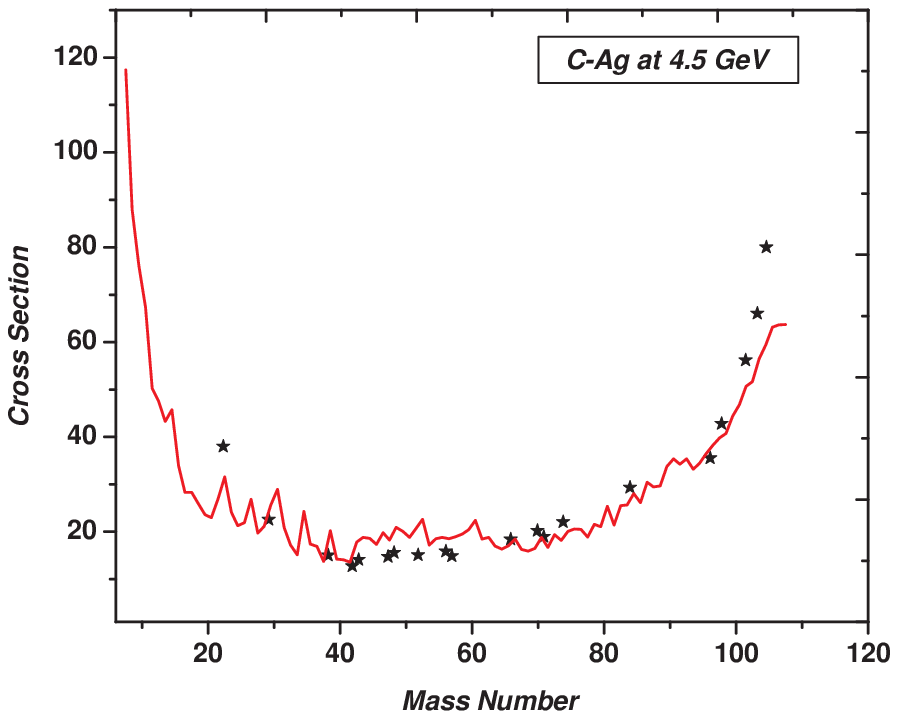}
\caption{Mass distribution of fragments in the reaction $^{12}$C +
$^{108}$Ag at 4.5 A GeV; $p_{bond}=0.72$. Data are taken from paper
\protect\cite{kozma}.}
\end{center}
\end{figure}

 As the bombarding energy increases, so does the contribution of
multifragmentation channels and, correspondingly, the contribution of
evaporation and fission channels decreases. This leads to filling of the
dips at both sides of central bell-shaped curve and a decrease in the
height of the right peak. As mentioned above, rising collision energy
leads to an increasing number of broken sites and bonds in the nuclear
lattices. This, in turn, results in increasing yield in multifragmentation
channels. This tendency is already evident in reaction Au + p at 0.8 A GeV
(Fig. 2).  The mass distributions of fragments in proton- and
carbon-induced reactions on Ag and Au at energies corresponding to the
limiting fragmentation regime are shown in Figures 8 -- 10.

\section{Conclusions}

A new version of the Modified Cascade Model for intermediate and high
energy nucleus-nucleus collisions including multifragmentation channels
has been developed. Colliding nuclei are represented as face--centered--
cubic lattices. Multifragmentation is calculated in the framework of
percolation theory with usage of a site--bond percolation model. This
version is able to reproduce reasonably well both spallation and
multifragmentation processes.

\section{Appendix: Hadronic Event Generator}

Monte Carlo simulation of inelastic events is performed in several steps.
The first step in the generation of an exclusive event is the evaluation
of the initial c.m. energy portion available for production of secondaries
\begin{equation}
W=\sum E_{i}=k\sqrt{s}
\end{equation}
where $E_{i}$ is the energy of the i--th particle (excluding leading
particles), $k$ is inelasticity. Fluctuation of the inelasticity from
event to event leads to the distribution $P(k)$. There are no comparable
theoretical methods for calculation of $P(k)$. It has been shown in Ref.
\cite{fowler} that one may fit the inelasticity distribution with a beta
distribution
\begin{equation}
P(k,s)=k^{a-1}(1-k)^{b-1}/B(a,b)
\end{equation}
\begin{equation}
B(a,b)=\Gamma (a)\Gamma (b)/\Gamma (a,b)
\end{equation}
\begin{equation}
<k(s)>=a/(a+b)
\end{equation}
where $\Gamma (a)$, $\Gamma (b)$ and $\Gamma (a,b)$ are gamma functions;
$s$ is the dependence of $P(k,s)$ and $<k(s)>$ is enclosed in parameters
$a$ and $b$. Up to the ISR energies one can neglect this $s$--dependence.
In the second step the energy $W$ is distributed between secondary
particles whose kinematical characteristics are generated in
correspondence to a cylindrical phase space model. Parameters of the
cylindrical phase space model are adjusted by comparing the results of
simulation of pion--nucleon and nucleon--nucleon interactions with
experimental data. The remaining part of c.m. energy $(1-k) \sqrt{s}$ is
distributed between remnants of the interacting particles (so--called
leading particles) according to the conservation of energy and momentum.
\begin{equation}
\overline{P}_{I}+\overline{P}_{II}=\sum \overline{P}_{i}
\end{equation}
\begin{equation}
E_{I}+E_{II}=(1-k)\sqrt{s}
\end{equation}
where $\overline{P}_{i}$ is the momentum of the $i$-th produced particle,
and $\overline{P}_{I}$, $\overline{P}_{II}$ and $E_{I},E_{II}$, are
momenta and energies of leading particles. Interacting nucleons (mesons)
can transform into nucleons (mesons) and s -- wave resonances ($\Delta $
-- isobars and $\rho $, $\omega $ -- mesons). Transition probabilities are
calculated with the use of a one--pion-exchange (OPE) model.

\end{document}